\documentclass[useAMS,usenatbib]{mn2e}
\usepackage{times}
\newcommand{\kms}{km~s$^{-1}$}
\usepackage{graphics,epsfig}

\title[The dwarf galaxy HIZSS003]{Resolving the mystery of the dwarf galaxy HIZSS003}
\author[Begum et al.]
{
Ayesha Begum$^{1}$\thanks{E-mail:ayesha@ncra.tifr.res.in},
Jayaram N. Chengalur$^{1}$, 
I. D. Karachentsev$^{2}$ and
M. E. Sharina$^{2}$
\\
\\
$^{1}$National Centre for Radio Astrophysics, Post Bag 3, Ganeshkhind, Pune 411 007, India\\
$^{2}$Special Astrophysical Observatory, Nizhnii Arkhys 369167, Russia\\
}
\begin{document}

\date{}

\pagerange{\pageref{firstpage}--\pageref{lastpage}} \pubyear{}

\maketitle

\label{firstpage}

\begin{abstract}

      The nearby galaxy HIZSS003 was recently discovered during a blind 
HI survey of the zone of avoidance (Henning et al. 2000). Follow up VLA as well 
as optical and near-IR imaging and spectroscopy (Massey et al. 2003; Silva et al. 2005)
confirm that it is a low metallicity dwarf irregular galaxy. However there were two 
puzzling aspects of the observations, (i)~current star formation, as traced by
H$\alpha$ emission, is confined to a small region at the edge of the VLA HI image
and (ii)~the metallicity of the older RGB stars is higher than that of the gas in
HII region. We present high spatial and velocity resolution Giant Meterwave
Radio Telescope (GMRT) observations that resolve these puzzles by showing that
HIZSS003 is actually a galaxy pair and that the HII region lies at the center
of a much smaller companion galaxy (HIZSS003B) to the main galaxy (HIZSS003A). 
The HI emission from these two galaxies overlaps in projection, but can be
separated in velocity space. HIZSS003B has an HI mass of $2.6 \times 10^6$~M$_\odot$,
and a highly disturbed velocity field. Since the velocity field is disturbed, an accurate
rotation curve cannot be derived, however, the indicative dynamical mass is
$\sim 5\times 10^7$~M$_\odot$. For the bigger galaxy HIZSS003A we derive
an HI mass of $1.4\times 10^7$~M$_\odot$. The velocity field
of this galaxy is quite regular and from its rotation curve we derive
a total dynamical mass of $\sim 6.5\times 10^8$~M$_\odot$.

\end{abstract}


\section{Introduction}
     
      The galaxy HIZSS003 was discovered in the course of a blind HI 21cm 
survey of the Zone of Avoidance (ZOA) (Henning et al. 1998, Henning et al. 2000). Its very low galactic 
latitude (b=0.09$^\circ$) made identification of the optical counterpart in the 
Palomar Sky Survey images difficult; however, the HI properties derived from the 
single dish data were consistent with it being a dwarf irregular galaxy. This was 
later confirmed by follow-up VLA~D array and optical observations (Massey et al., 2003). 
The VLA map showed two peaks in the HI distribution, a  resolved peak at the center 
of the galaxy and an unresolved secondary peak close to the edge of the HI distribution. 
While broadband BVRI  imaging failed to detect any stars in the galaxy, 
narrowband H$\alpha$ imaging detected an HII region spatially coincident  with the 
unresolved secondary HI peak in the VLA map.  Spectroscopy of  the HII region 
confirmed that its radial velocity agreed with that of the  HI emission.  Although 
the identification of an HII region strengthens the conclusion that HIZSS003 is 
a dwarf irregular galaxy,  it is puzzling that the current star formation should 
be concentrated at the edge of the HI disk. VLT near-IR images as well as  MMT 
spectroscopic data of the HII region were presented by Silva et al. (2005). The 
near-IR images revealed a resolved stellar population and allowed the distance 
to the galaxy to  be derived based on the K magnitude of the tip of the red 
giant branch (TRGB). The derived distance of $1.69 \pm 0.07$~Mpc agrees well 
with the previous estimate of 1.8~Mpc by Massey~et al.~(2003)  (based on the 
assumption that HIZSS003 has zero peculiar velocity with respect to the local 
group centroid). 

     From their spectroscopic data Silva et al. (2005) estimate the metallicity 
([O/H]) of the HII region in HIZSS003 to be $\sim -0.9$, comparable to that of
other nearby metal poor irregular galaxies. On the other hand, the metallicity
([Fe/H]) of the RGB stars (as derived from their colours) is somewhat higher, 
viz. $-0.5 \pm 0.1$. That the older RGB stars in the galaxy are more metal rich 
than the gas associated with on going star formation in the HII region is puzzling. 
Silva et al. (2005) speculate that the lower metallicity of the HII region is caused
by low metallicity  gas that is falling into HIZSS003 for the first time.  Here 
we present high resolution GMRT HI images of HIZSS003, which resolve the puzzles 
of the HII region location and metallicity by showing that HIZSS003 is in fact 
a galaxy pair, with the HII region being located at the center of a much smaller 
companion to the main galaxy.  Throughout this paper we adopt the Silva et al. (2005)
TRGB distance estimate of 1.69~Mpc for HIZSS003.

\section[]{Observations and data analysis }
\label{sec:obs}

    The GMRT (Swarup et al. 1991) observations of HIZSS003 (RA (2000): 07$^h$00$^m$29.3$^s$,
DEC(2000): $-{04}^{\circ} 12^\prime 30^{\prime\prime}$ )  were conducted on
23 Aug 2004. An observing bandwidth of 1 MHz centered at 1419.1 MHz (which
corresponds to a heliocentric velocity of 290 \kms) was used . The band
was divided into 128 spectral channels, giving a channel spacing of 1.65 \kms.
Flux calibration was done using scans on the standard calibrators
3C147 and 3C286, which were observed at the start and end of the observing run.
Phase calibration was done using 0744-064, which was observed once every 40
minutes. Bandpass calibration was done in the standard way using 3C286. The
total on-source time was $\sim$ 4 hours.

     The data was reduced using standard tasks in classic AIPS.  The GMRT  has a hybrid 
configuration which simultaneously provides both high angular resolution ($\sim 2^{''}$ if 
one uses baselines between the arm antennas) as well as sensitivity to extended emission
(from baselines between the antennas in the central array). Data cubes were therefore  
made at various resolutions including 42$^{''}\times 39^{''}$, 28$^{''}\times 26^{''}$, 
23$^{''}\times 18^{''}$, 18$^{''}\times 11^{''}$, 8$^{''}\times 6^{''}$ and $4^{''}\times3^{''}$,
using uniform weighting. RMS noise per channel for these resolution is
2.2~mJy, 2.0 mJy, 1.8 mJy, 1.6 mJy, 1.4 mJy and 1.2 mJy respectively.
All the data cubes, except  8$^{''}\times 6^{''}$ and $4^{''}\times3^{''}$,
were deconvolved using the task IMAGR. For the two highest resolution
data cubes, the signal to noise ratio was too low for CLEAN to work
reliably. A continuum image was made using the average of the line free 
channels. No continuum was detected from the galaxy to a $3\sigma$ flux limit  
of 1.0~mJy/beam (for a beam size of $28^{''}\times26^{''}$). A high resolution 
continuum map ($4^{''}\times3^{''}$ resolution) was also made to search for any 
compact continuum sources. The only continuum source of note detected is NVSS 
J070023-041255. The HI column density (as derived from the $42^{''}\times 39^{''}$ 
resolution image) along the line of sight to this source  is 5.7$\times~10^{20}$ 
atoms cm$^{-2}$.  A search for HI absorption, in the direction of this source, 
gave negative results at all resolutions. The implied lower limit on the spin 
temperature of the gas (assuming a  velocity width of 10 \kms) is 723 K. For 
reference we note that NVSS J070023-041255 lies  towards the source that we 
call HIZSS003B below, and that it is close to, but not coincident with
the HII region detected by Massey et al. (2003). It appears likely that it is
a background source, with no connection to the HI emission. The continuum source 
was subtracted using the task UVSUB. 

\section[]{Results and Discussion}
\label{sec:res}

\begin{figure*}
\psfig{file=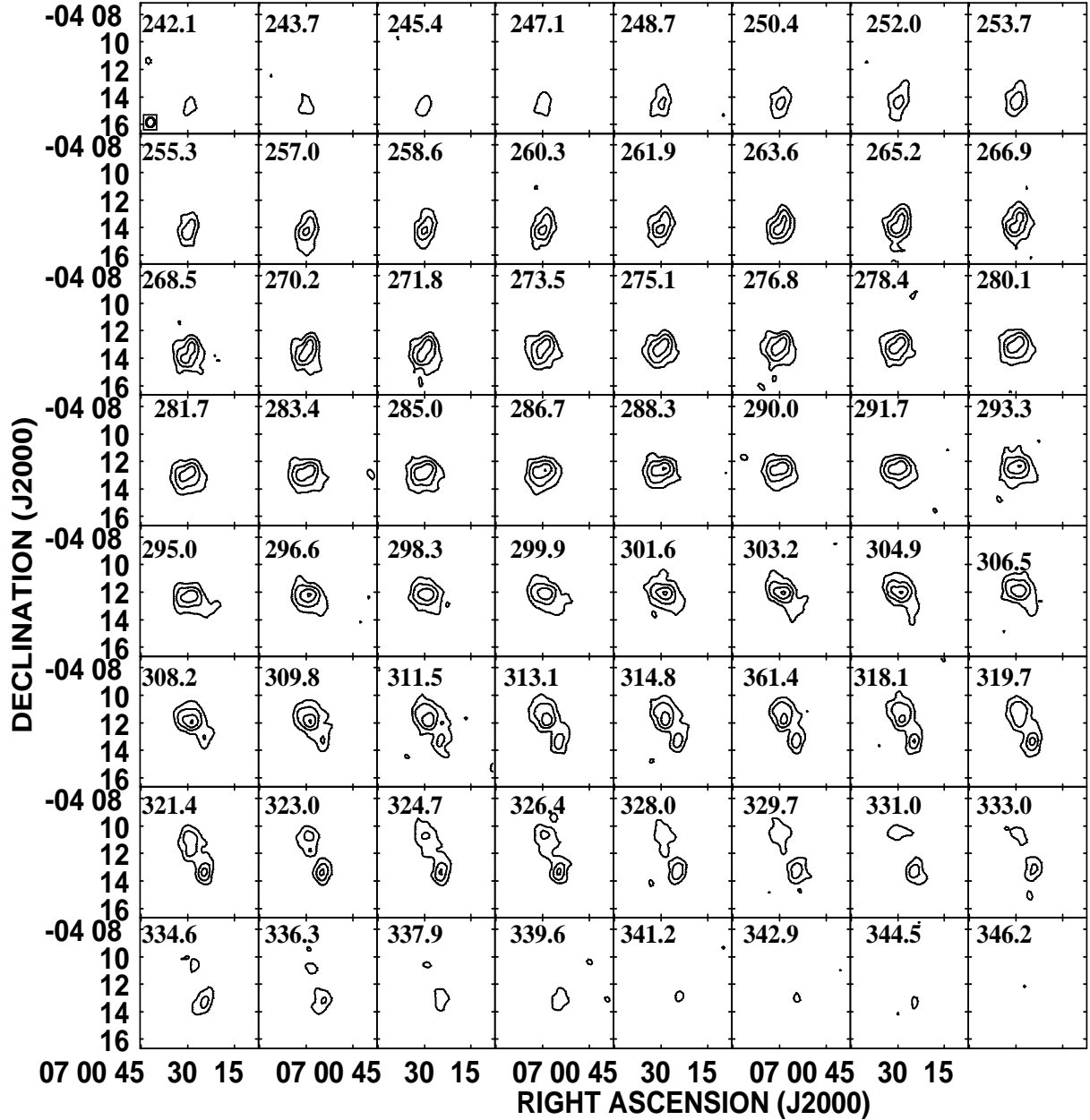,width=6.6truein,height=7.5truein}
\caption{HI channel maps of HIZSS003 at 42$^{''}\times39^{''}$ resolution.
The contour levels are 7.5, 22.5, 45 and  75 mJy \kms. The heliocentric 
velocity in \kms is marked on the upper left-hand corner of every pane.  
The channel spacing is 1.65 \kms. 
}
\label{fig:cube}
\end{figure*}

\begin{figure*}
\psfig{file=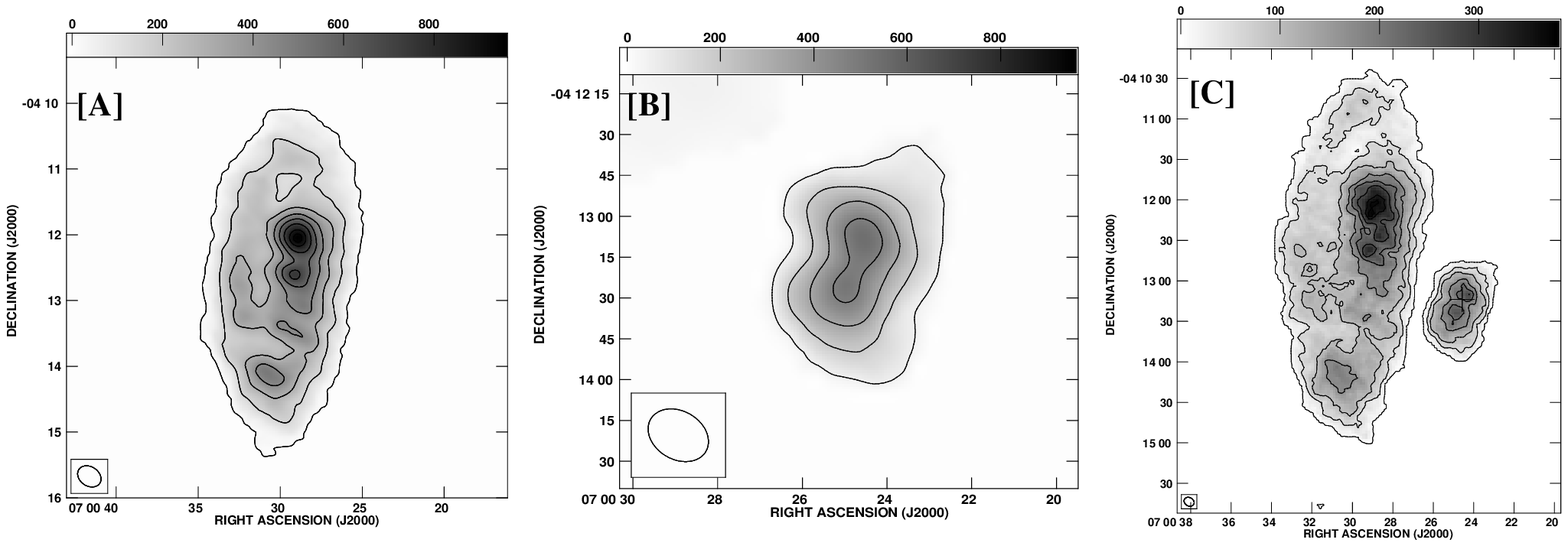,width=7.0truein}
\caption{ Integrated HI emission maps (grey scales and contours) of HIZSS003A (panel {\bf{[A]}})
	and HIZSS003B (panel {\bf{[B]}}) at 23$^{''}\times18^{''}$ resolution. 
        The contour levels are 0.2, 3.9, 7.8, 11.6, 15.4, 19.1, 31.4 and 24.9 $\times10^{20}$ atoms cm$^{-2}$.
        The angular scale for panel {\bf{[B]}} has been expanded for clarity.	
          {\bf{[C]}} Integrated HI emission map of HIZSS003 system (grey scale and contours)
          at 8$^{''}\times6^{''}$ resolution. The contour levels are 
	 0.03, 0.05, 0.10, 0.17, 0.21, 0.27, 0.34 and 0.36 Jy/beam \kms. The location of HII region
         in HIZSS003B is marked by a cross.
}
\label{fig:mom0}
\end{figure*}

    Channel maps of the HI emission at a resolution of $42^{''}\times39^{''}$
are shown in Fig.~\ref{fig:cube}. HI emission is spread over 63 channels and
consists of two distinct sources, one spanning 59 channels and the other
spanning 26  channels. At this spatial resolution some channels show HI emission 
connecting the two sources, however it is not clear whether this is due to beam 
smearing. A HI feature connecting the two sources is also seen in
28$^{''}\times 26^{''}$ and 23$^{''}\times 18^{''}$ resolution data cubes.
However, at higher resolutions no such connecting emission is seen in 
the channel maps. Further, as discussed  in more detail below, the velocity 
field of the bigger source does not appear to be particularly disturbed, and 
neither source shows signs of two armed tidal distortions. It is possible
that the connecting emission seen in Fig.~\ref{fig:cube} is due
to beam smearing. In order to disentangle the HI emission one spectral cube was
made for each galaxy in which emission from the other galaxy was blanked out. 
In the case of channel maps which showed  connecting HI, 
the blanking was done midway between the two sources. The 
Fig.~\ref{fig:mom0}[A]\&[B] show HI images of the two sources at 
23$^{''}\times  18^{''}$ resolution made from these blanked cubes. In the 
rest of the paper, we refer to the the bigger (eastern) galaxy as HIZSS003A and 
the smaller  (western) one as HIZSS003B. The entire HI distribution will be 
referred as  the ``HIZSS003 system". The sources HIZSS003A and HIZSS003B 
correspond to the main HI peak and the secondary unresolved peak in the 
VLA map of Massey et al. (2003). The combination of poor spatial 
($\sim 60^{\prime\prime}$) and velocity ($\sim$10 \kms) resolution 
of the VLA observations prevented Massey et al. (2003) from separating
the two galaxies, although the near-IR VLT images do show two separate
stellar concentrations, i.e. one for each galaxy. Fig.~\ref{fig:mom0}[C] 
shows the high resolution HI map 
(8$''\times6''$ resolution) of HIZSS003. The more diffuse emission is 
resolved out, and the remaining emission from the two galaxies can be 
disentangled without having to resort to channel by channel blanking. 
As can be seen in the Fig.~\ref{fig:mom0}, the HI distribution in both 
the galaxies is clumpy,  with three main peaks  seen in the HI distribution 
of HIZSS003A, whereas the HI distribution of HIZSS003B is resolved into two peaks.
No signature of tidal interaction is evident  in the HI distribution of either 
galaxy. The HII region detected in the HIZSS003 system is located close
to one of the peaks of the HI distribution in HIZSS003B (shown by a cross
in Fig.~\ref{fig:mom0}[C]). The H$\alpha$ emission is approximately
aligned with the HI contours of the galaxy (i.e. from northwest to southeast),
and its heliocentric velocity (335$\pm$15 \kms -- Massey et al. 2003), matches
within the error bars with the systemic velocity of 322.6 $\pm$ 1.4 \kms 
for  HIZSS003B  derived from the HI global profile (see below).

 Fig.~\ref{fig:spectra} shows the global HI emission profiles of the two galaxies
obtained from the $42{''}\times39^{''}$ resolution data cubes. As discussed above, emission 
from one galaxy was blanked before obtaining the HI profile for the other one. 
Gaussian fits to the HI profiles give systemic velocities of 288.0$\pm$2.5 \kms
and 322.6 $\pm$ 1.4 \kms for HIZSS003A and HIZSS003B respectively. The corresponding 
velocity widths at 50\% of peak emission are 55 \kms and 28 \kms, while the integrated 
fluxes are $20.9\pm2.1$ ~Jy~\kms and $3.8\pm0.3$~Jy~\kms. The HI masses corresponding
to these integrated flues are 1.4 $\times10^7 \rm{M_\odot}$ and 
2.6 $\times10^6 \rm{M_\odot}$. The combined flux of both galaxies is
24.7~Jy~\kms, which is in excellent agreement with the value of 24.9~Jy~\kms
obtained from the VLA observations by Massey et al. (2003). However both these
values are somewhat lower than the flux integral of $\sim$32 Jy~\kms, estimated from 
the single dish observations by Henning et al. (2000). The HI diameter of the two galaxies,
measured at a level of $\sim10^{19}$ atoms cm$^{-2}$ (from the 42${''}\times39{''}$ 
images) are $\sim6.5^\prime$ (3.2 kpc) and $\sim3^\prime$ (1.5 kpc).

Fig.~\ref{fig:mom1}[A]\&[B] show the velocity fields of the two galaxies 
derived from the moment analysis of 28$''\times26''$ resolution 
data cube. The velocity field of HIZSS003A (Fig.~\ref{fig:mom1}[A]) is regular
and a large scale velocity gradient, consistent with systematic rotation,
is seen across the galaxy. The velocity field is also mildly lopsided $-$ 
the isovelocity contours in the southern half of the galaxy are more curved 
than the northern half. Kinks are seen in the eastern isovelocity contours,
close to the location of HIZSS003B. These kinks are more prominent in 
the higher resolution velocity fields (not shown).

   Rotation curves of HIZSS003A were derived using 42$^{''}\times 39^{''}$, 
28$^{''}\times 26^{''}$, 23$^{''}\times 18^{''}$ and 18$^{''}\times 11^{''}$ 
resolution velocity fields, using tilted ring fits. The center and systemic 
velocity for the galaxy obtained from a global fit to the various resolution 
velocity fields matched within the error bars; the systemic velocity of 
291.0$\pm$1.0 \kms also matched  with the value obtained from the global HI 
profile of the galaxy. Keeping the center and systemic velocity fixed, we fitted
for the inclination and position angle (PA) in each ring. For all resolution
velocity fields, the PA was found to vary from $\sim -4^\circ~\rm{to}~4^\circ$ and
the inclination varied from $\sim 70^\circ~\rm{to}~55^\circ$.
Keeping the PA  and inclination fixed to 3$^\circ$ and  65$^\circ$ in the 
inner regions (upto 90$^{\prime\prime}$) and 0$^\circ$ and  60$^\circ$ in the 
outer regions respectively, the rotation curves at various resolutions were derived. Fig.
~\ref{fig:mom1}[C] shows the rotation curve of the galaxy derived at various 
resolutions $-$ as can be seen, they match within the 
errorbars. The solid  line show the final adopted rotation curve. The total 
dynamical mass of HIZSS003A (at the last measured point of the rotation curve)
is found to be $6.5\times 10^8 \rm{M_\odot}$. 

The velocity field of HIZSS003B (Fig.~\ref{fig:mom1}[B]) shows a large scale
gradient in the southeast-northwest direction with a magnitude of $\sim$ 5 \kms kpc$^{-1}$.
This gradient is aligned along the direction of elongation of the HI contours and also 
with the HII region in the galaxy (Fig.~\ref{fig:mom0}[B]). However, the observed velocity
pattern is clearly not rotation since the velocity gradient is not monotonic. Both ends 
of the galaxy are at a higher velocity than the central region. Similar kinematics 
are seen in other very low mass dwarf galaxies e.g. Sag DIG and LGS 3 (Young \& Lo 1997).
The disturbed kinematics may be due to a combination of tidal perturbation from
the companion galaxy and energy input from the ongoing star formation (see e.g.
GR8, Begum \& Chengalur 2003).  The inclination of HIZSS003B derived from the ellipse 
fit to the HI distribution (assuming an intrinsic axial ratio q$_0$=0.25) is $\sim50^\circ$,
which means that the upper limit to the inclination corrected rotation
velocity is $5/\sin(50) \sim 6.5$~\kms. Assuming  that the gas has a velocity dispersion ($\sigma$) 
of 8 \kms, (which is a typical value for low mass dwarf  irregular galaxies, e.g. 
Lake et al. 1990, Begum et al. 2003),  implies that the systematic
rotation, if any, in the galaxy is smaller than the velocity dispersion.
Given the lack of any systematic rotation, it is difficult to accurately determine 
the total dynamical mass for the galaxy. From the virial theorem, assuming HI distribution 
to be spherical with an isotropic velocity dispersion and negligible rotation, the indicative 
mass is (Hoffman et al. 1996)
\begin{equation}
\rm{M_{VT}=  {5~r_H \times \sigma^2 \over  G}}
\end{equation}

 Assuming $\sigma$ of 8 \kms and taking the diameter of
the galaxy $\sim $ 1.5 kpc, gives a total mass of HIZSS003B to be 
$\sim 5.3\times10^7 \rm{M_\odot}$. For the entire HIZSS003 system, if we assume the two
galaxies to be in a bound circular orbit, then the indicative orbital mass
is 
\begin{equation}
{\rm M_{orb} = {32\over 3\pi G} r_p {\Delta V}^2}
\end{equation}

where $r_p$ is the projected separation and $\Delta V$ the radial velocity difference
(Karachentsev et al. 2002). For a projected separation of $\sim 0.7$~kpc and a velocity
difference of $\sim 34.6$~\kms, the indicative orbital mass is $\sim 6.7\times 10^8$~M$_\odot$,
in good agreement with the total mass derived from the internal kinematics.

\begin{figure}
\psfig{file=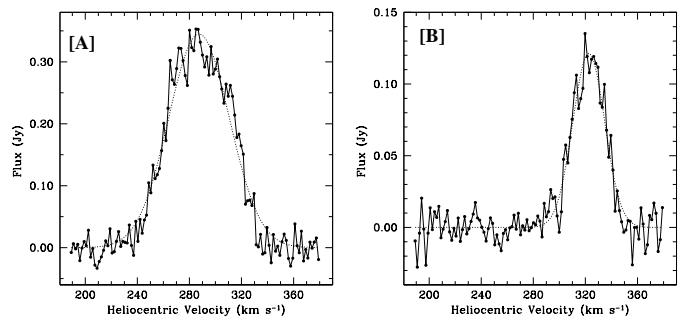,width=3.6truein}
\caption{Global HI emission profiles for HIZSS003A (panel {\bf{[A]}}) and HIZSS003B (panel {\bf{[B]}})
         obtained from the 42$^{''}\times39^{''}$ data cube. The dashed lines show  gaussian fits 
         to the profiles. 
}
\label{fig:spectra}
\end{figure}

\begin{figure*}
\psfig{file=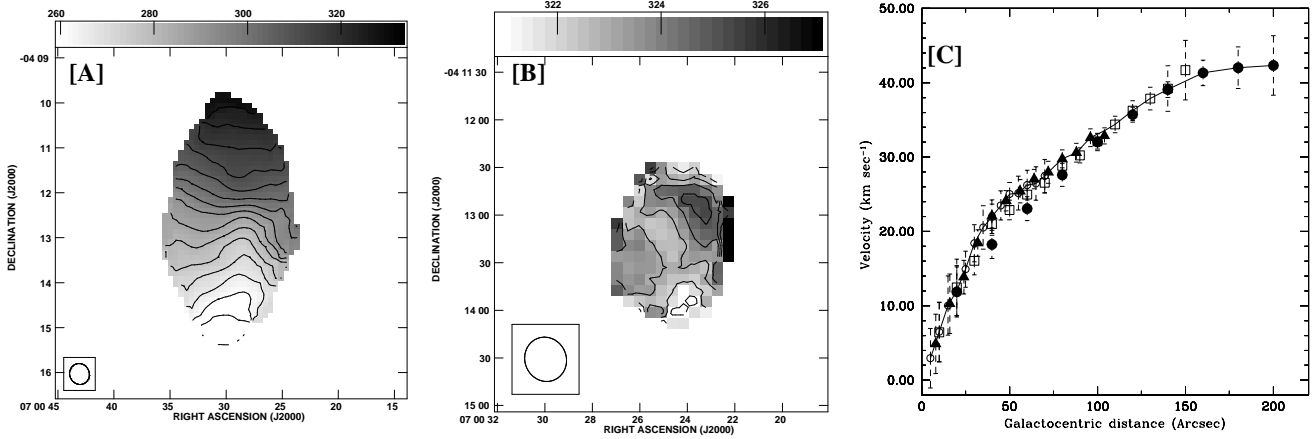,width=7.0truein}
\caption{{\bf{[A]}}The HI velocity
        field of HIZSS003A at 28$^{''}\times 26^{''}$ resolution. The
        contours are in the steps of 5 \kms and range from 251.0~\kms
        to 326.0~\kms.
          {\bf{[B]}}The HI velocity
        field of HIZSS003B at 28$^{''}\times 26^{''}$ resolution. The
        contours are in the steps of 1 \kms and range from 321.0~\kms
        to 326.0~\kms. The angular scale for this figure has been expanded for clarity.
          {\bf{[C]}}The rotation curve for   HIZSS003A derived from the
        intensity weighted velocity field at 42$^{''}\times 39^{''}$,
        28$^{''}\times 26^{''}$,  23$^{''}\times18^{''}$ and 18$^{''}\times 11^{''}$
        resolution (shown as filled circles, open squares, filled triangles and open circles
        respectively). The solid line shows the adopted rotation curve.
}
\label{fig:mom1}
\end{figure*}

    Silva et al. (2005) highlight a puzzle regarding the  metallicity 
of HIZSS003. The metallicity of HIZSS003 system calculated from the 
younger HII region is smaller than that  estimated from the color
of the older red giant branch stars. Given that the bulk of the stars 
are associated with the bigger galaxy but that the HII region is 
in the smaller galaxy, the inconsistency in the derived metallicities 
is not surprising.  The low metallicity of the gas in the 
HII region of the smaller galaxy HIZSS003B is also qualitatively consistent 
with what one would expect from the metallicity-luminosity relation.
Further, going by the stars identified as belonging to the HII region 
(Fig. 6 of Silva et al. 2005)  there is a trend for these stars 
to have a slightly smaller J-K color  (consistent with a lower metallicity; 
Valenti et al. 2004) than the median color of all stars identified as 
belonging to HIZSS003.  The HIZSS003 puzzle thus seems (at a qualitative level 
at least) resolved. However, as is well known, an observed colour difference 
cannot be uniquely ascribed to a difference in the metallicity, but could
also be due to a difference in the age of the stars ("age-metallicity" degeneracy)
or from  a temperature difference. A more quantitative  consistency check will have 
to await a detailed reanalysis of the near-IR data.

\section*{Acknowledgments}

        The observations presented in this paper were made
with the Giant Metrewave Radio Telescope (GMRT). The GMRT is operated
by the National Center for Radio Astrophysics of the Tata Institute
of Fundamental Research.

\end{document}